# Abrupt orthorhombic relaxation in compressively strained ultra-thin SrRuO$_3$ films

Prahald Siwakoti[1], Zhen Wang[1,2], Mohammad Saghayezhian[1], David Howe[1], Zeeshan Ali[1], Yimei Zhu[2], Jiandi Zhang[1]

[1]Department of Physics and Astronomy, Louisiana State University, Baton Rouge, LA 70803, USA

[2]Department of Energy Science and Technology, Brookhaven National Laboratory, Upton, NY 11973, USA

**Abstract**

Lattice structure can dictate electronic and magnetic properties of a material. Especially, reconstruction at a surface or heterointerface can create properties that are fundamentally different from those of the corresponding bulk material. We have investigated the lattice structure on the surface and in the thin films of epitaxial SrRuO$_3$ with the film thickness up to 22 pseudo-cubic unit cells (u.c.), using the combination of surface sensitive low energy electron diffraction and bulk sensitive scanning transmission electron microscopy. Our analysis indicates that, in contrast to many perovskite oxides, the RuO$_6$ tilt and rotational distortions appear even in single unit cell SrRuO$_3$ thin films on cubic SrTiO$_3$, while the full relaxation to the bulk-like orthorhombic structure takes 3-4 u.c. from the interface for thicker films. Yet the TiO$_6$ octahedra of the substrate near the interface with SrRuO$_3$ films show no sign of distortion, unlike those near the interface with CaRuO$_3$ films. Two orthogonal in-plane rotated structural domains are identified. These structural distortions are essential for the nature of the thickness dependent transport and magnetism in ultrathin films.

Tailoring the lattice mismatch strain in epitaxial films and heterostructures has been used as a tool to stabilize novel phases otherwise non-existent in the bulk [1], to enhance the physical properties such as superconductivity [2,3], ferroelectricity [4], and ferromagnetism [5,6], as well as to manipulate magnetic anisotropy [7], and metal-insulator transitions [8–11]. The effect of strain mismatch on the electronic and magnetic properties of SrRuO$_3$ (SRO) thin films has been extensively studied for decades [12–14]. Bulk SRO at room temperature crystallizes in orthorhombic structure with *Pbnm* (No. 62) symmetry with lattice parameters $a$ = 5.5670 Å, $b$ = 5.5304 Å and $c$ = 7.8446 Å. An orthorhombic unit cell is produced by rotations of RuO$_6$ octahedra from the ideal cubic structure. Such rotations in SRO produce a distorted pseudo-cubic perovskite structure with a pseudo-cubic lattice constant of 3.93 Å [15,16]. In Glazer notation [17], it can be described with $a^+b^-b^-$ tilt system (#10) which is defined by in-phase rotation of the neighboring octahedra about the [100]$_p$ pseudo-cubic axis (hereafter denoted with a subscript "p") and mutually equivalent out-of-phase rotation about [010]$_p$ and [001]$_p$ axes, respectively. When grown as thin films, the interfacial octahedral mismatch between substrate and film creates a coupling effect. The accommodation of the octahedral symmetry mismatch is achieved through rigid rotations and/or deformation of octahedra near the interface. The spatial extent of this accommodation near the interface is found to be dependent on the rigidity of the octahedral network. He *et al* [18] studied, by density functional theory, the structural effect of interfacial symmetry mismatch accommodation on two model interfaces of SRO and $La_{0.75}Sr_{0.25}MnO_3$ (LSMO) thin films grown on $SrTiO_3$ (STO) substrate. They found that in the case of SRO, the octahedra near the interface deform rapidly to the bulk value in fewer than three $RuO_2$ layers. For LSMO, on the other hand the octahedral deformation takes place for several layers into the thin film. It was also found that the physical properties such as large perpendicular magnetic anisotropy are more sensitive to octahedral tilt angles in SRO compared to LSMO, due primarily to the strong hybridization of Ru *d* and oxygen *p* orbitals [18]. In addition to the distortion on the film side, lattice distortions near the interface of cubic STO side can also happen. It is observed the LSMO thin film can induce polar distortion [19] as well as tilt distortion [20] of $TiO_6$ near the interface of STO.

Chang *et al* [21], in their X-ray reciprocal space map study, found that the orthorhombicity in SRO films disappears below a thickness of ~18 unit cells. They concluded that, by examining certain half-order Bragg reflection peak, the orthorhombicity is caused by the presence of octahedral tilts about the $[001]_p$ axis. While the presence of such half-order peaks can be a proof of the lowering of the symmetry due to tilts/rotation, the observation of these fractional peaks can be obscured due to weak signal intensity from the ultrathin films. Scanning transmission electron microscopy (STEM) is another valuable tool to study atomically resolved structure of thin films. However, for ultrathin films, surface effects can significantly affect the observed properties [22]. Capping with such as STO layer to protect the surface can suppress the octahedral tilts/rotations of the underlying film [23], thus bringing extrinsic effects to the system.

In order to avoid the above mentioned difficulties, we have combined *in-situ* low energy electron diffraction (LEED) characterization of films with varying thicknesses with *ex-situ* STEM studies to investigate the symmetry of the structure of films down to single u.c. in thickness. LEED imaging is sensitive to changes in the symmetry near the surface of a film or substrate. Therefore, it can be used to monitor structural distortions that are otherwise difficult to observe through STEM and X-ray diffraction. One advantage of LEED is that it can be performed on a pristine film surface without further processing of any kind. A bulk truncated cubic perovskite material surface [STO (001) surface, for example] has a square symmetry. Therefore, the diffraction pattern consists of the Bragg spots that lie on a square 2D reciprocal lattice. If thin film grown on this substrate follows the same symmetry of the substrate surface, we expect a $p(1\times1)$ LEED pattern. Any change in the symmetry of the grown film with respect to the cubic substrate structure would be reflected in the LEED pattern. In this study, we observe previously not reported $(\sqrt{2} \times \sqrt{2})R45°$ surface symmetry in ultra-thin $SrRuO_3$ films. We ascribe this structure to the reduced symmetry of the surface caused by orthorhombic distortion of the film, which is further confirmed by STEM. These orthorhombic distortions appear even in the film with a single layer of SRO, in general agreement with the theoretical study of He et al [18]. Furthermore, the interface layers of STO show no sign of structural distortion even with such an abrupt orthorhombic relaxation of SRO film as determined by STEM. Our study demonstrated that a combination of *in-situ* LEED and *ex-situ* STEM measurements could be powerful to identify the microscopic lattice structure at the interface and in ultrathin films.

The thin film growth was achieved with pulsed laser deposition (PLD). A stoichiometric SRO target was illuminated with a KrF excimer laser ($\lambda = 248$ nm) at a repetition rate of 10 Hz and a laser energy of 350 mJ. The oxygen partial pressure of 60 millitorr (mTorr) was obtained with a mixture of 99% $O_2$ and 1% $O_3$ and maintained during growth while the substrate temperature was kept at 700 °C. The growth was monitored via reflection high-energy electron diffraction (RHEED) intensity oscillations. The intensity of RHEED diffraction spot oscillates for a few unit cells of deposition and eventually saturates, as shown in Fig. 1(a), suggesting a typical change in growth mode [24] of SRO from layer-by-layer mode to step flow mode. Due to the volatile nature of $RuO_2$, SrO termination is preferred and corresponds to the peak in the RHEED intensity oscillations [25]. The period of the first peak in the RHEED intensity is about 1.5 times the average period of subsequent peaks suggesting that a termination conversion happens during the initial growth of SRO films on $TiO_2$ terminated substrate surface.

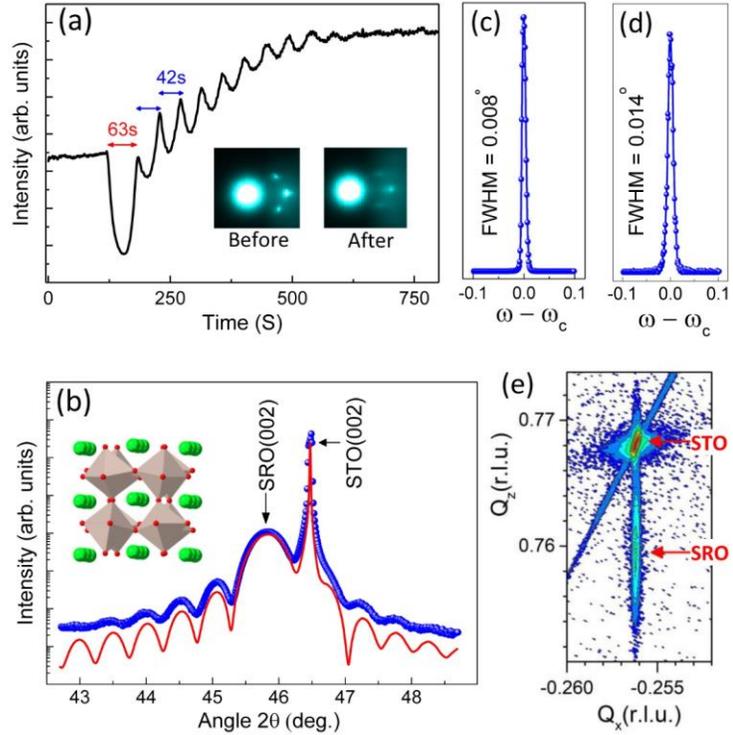

**Fig.1** (a) RHEED oscillations of the initial growth of SRO films on STO(001) surface. The inset shows the diffraction pattern before and after film growth. (b) Coupled X-ray diffraction scan of a ~47 unit-cell thick SRO film. The Laue oscillations are fitted and presented in the red curve. Inset shows a schematic orthorhombic structure of bulk SRO. (c-d) show rocking curves of the substrate and film peak, respectively. e) Reciprocal Space Mapping about the (-103) Bragg diffraction spot. The substrate peak and the film peak are marked.

Figure 1(b) shows a symmetric ($2\theta - \omega$) X-ray Diffraction (XRD) scan about the pseudo-cubic (002) Bragg peak of SRO thin film grown on STO(001) substrate. Laue oscillations are observed, indicating smooth interface and surface. These Laue fringes along with the main reflection peak can be fitted to estimate the film thickness. The fit alongside the experimental data in Fig. 1(b) is obtained by using a modified kinematical theory of diffraction, wherein absorption effects are taken into account by introducing an attenuation factor [26]. The out-of-the-plane lattice constant increased to 3.955 Å from the bulk pseudo-cubic value of 3.926 Å [15]. Such a lattice expansion is expected because of the compressive strain from the substrate. Figures 1(c-d) show the rocking curve of the substrate and the film, respectively. The full width at half maximum (FWHM) of the film rocking curve (0.014 Å) is close to the FWHM of the substrate rocking curve (0.008 Å), suggesting that the crystallinity of the thin films is comparable to that of the substrate. Figure 1(e) shows the Reciprocal Space Map (RSM) about the (-103) asymmetric diffraction peak. The substrate peak and the film (-103) peak are labelled. The vertical $Q_z$ axis is parallel to the out-of-plane direction of the film. The horizontal direction ($Q_x$ axis) is the in-plane direction. The film and substrate peaks line up in the perpendicular direction suggesting the film is highly strained in-plane to the substrate.

Once the growth conditions for high-quality epitaxial thin film growth of SRO were optimized, we grew the films with different thicknesses, including ones with extremely low thickness, to probe their structure with LEED. The LEED pattern of a bulk truncated (001) surface of STO substrate is shown in Fig. 2(a). The dashed red square is the $p(1\times1)$ unit-cell. Fig. 2(b) displays the LEED pattern of a 5 u.c. SRO film grown at an oxygen partial pressure $P_O = 60$ mTorr, at a beam energy of 74 eV. In addition to the integer spots, the unit cell for which are denoted by the dashed red square, we also observe the spots corresponding to the $(\sqrt{2} \times \sqrt{2})R45°$ unit cell as shown by the solid blue square. Subsequent annealing at 630°C and $P_O = 100$ mTorr for 30 minutes [Fig. 2(c)] did not remove the $(\sqrt{2} \times \sqrt{2})R45°$ spots as the first order spots at $(\pm\frac{1}{2}, \pm\frac{1}{2})$ are clearly observed, though the background of the LEED pattern is slightly enhanced. The intensities of the fractional spots are comparable with that of the integer spots, suggesting that the observed $(\sqrt{2} \times \sqrt{2})R45°$ pattern is due to lattice distortion result of whole film rather than a surface reconstruction. More annealing at the same temperature but higher oxygen partial pressure ($P_O = 150$ mTorr for 30 more minutes) [Fig. 2(d)] did not change the $(\sqrt{2} \times \sqrt{2})R45°$ LEED pattern, except an increase of background which indicates an enhancement of surface disorder likely due to extra disordered oxygen adatoms. However, upon subsequent vacuum annealing for 30 minutes, the background decreased, and the high-order fractional spots became more apparent, as shown in Fig. 2(e). The vacuum annealing removes the surface disorder such that the pattern in Fig. 2(e) resembles the pattern of as-grown film in Fig. 2(b). Surprisingly, the LEED pattern of a 1 u.c. SRO film displayed in Fig. 2(f) has the same symmetry as the 5 u.c. film, though high-order fractional spots are relative weak. This clearly indicates that a distortion from cubic perovskite structure begins with a monolayer of SRO film.

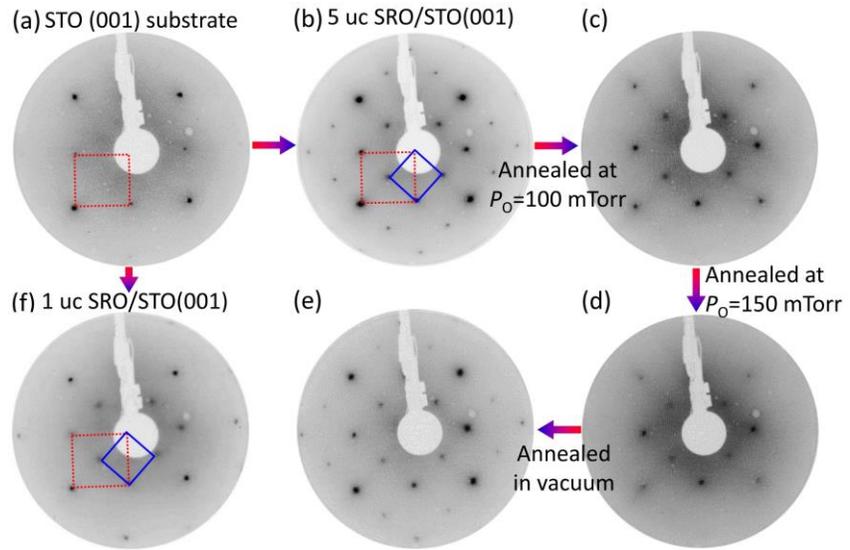

**Fig. 2** LEED images with a beam energy of 74 eV: (a) (1×1) Bulk truncated STO (001) surface overlapped with a red square for surface unit cell. (b) A 5 u.c. SRO film on STO (001) showing (√2×√2)R45° reconstructed unit cell (blue square) with respect to the substrate one. (c-e) The same sample under various annealing conditions. (f) A 1 u.c. SRO film.

We next consider the possibility of ordered oxygen vacancies or overlayer at the surface. For oxygen vacancies to create a $(\sqrt{2} \times \sqrt{2})R45°$ pattern, half the oxygen atoms of the top SrO layer need to be removed. Vacancies are generally not expected owing to the ionic character of SrO bonds. DFT studies [27] have predicted a rather high energy cost (~7.98 eV) of forming a single oxygen vacancy. Sr vacancies were also found to be equally energetically costly (7.19 eV/vacancy). A single oxygen adatom was found to adsorb favorably on a SrO terminated surface midway between the SrO oxygen atoms similar to that observed on the surface of manganites [28]. However, such scenario is in contradiction of our experimental observation. We did not observe any change in LEED pattern by vacuum annealing at 630 °C for 30 minutes (as shown in Fig. 2(e)) but rather an improved intensity due to decreasing background. Oxygen adatom-induced additional reconstructions have been reported in the literature for thicker SRO films. Shin *et al* [29] reported a $2 \times 2$ surface structure doubling with LEED and local domains of $1 \times 2$ and $2 \times 1$ symmetry with STM. However, it is worth mentioning that their samples were larger than 10 u.c. in thickness and were post annealed at very high oxygen pressure ($P_O$ = 1 Torr) for 90 minutes at 450°C. The robustness of these $(\sqrt{2} \times \sqrt{2})R45°$ spots under vacuum as well as oxygen annealing suggest that these are not likely a result of ordered oxygen vacancies and/or overlayers.

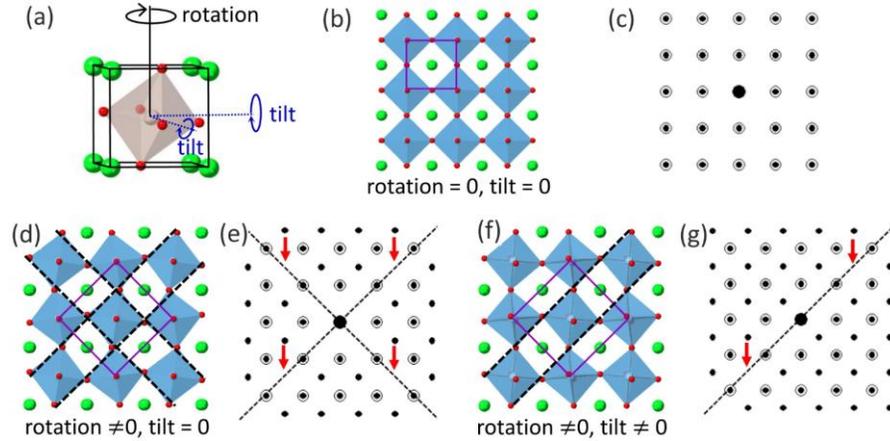

**Fig. 3** Simulated LEED diffraction patterns for different surface symmetry. (a) A schematic showing $RuO_6$ tilt and rotations. (b-c) top view of a square lattice with no $RuO_6$ rotation or tilt and the corresponding diffraction pattern. The integer spots due to a square lattice are shown as a dot with a concentric circle. (d-e) square lattice where $RuO_6$ octahedra are rotated but not tilted and the corresponding diffraction pattern. Fractional spots are shown as solid dots. Glide lines are shown with dashed lines. Arrows indicate the position of missing fractional spots. (f-g) Lattice with both $RuO_6$ tilts and rotation as well as the corresponding diffraction pattern. Only one glide line survives as shown with dashed lines.

With the suggestion of a pristine and stoichiometric thin film, we arrive at structural models for the surface with various configurations of tilts and rotations, as shown in Fig. 3. To describe the orthorhombic distortions from cubic structure, we define RuO$_6$ "*rotation*" as the rotation of the octahedra about the axis along the surface normal ([001] direction of the substrate) and "*tilt*" as the RuO$_6$ rotation about an axis in the film plane, either [100] and/or [010] direction of the substrate, as shown in Fig. 3(a). If the film grows on the cubic substrate without any tilt and rotation [Fig. 3(b)], the diffraction pattern would be expected to be $p(1 \times 1)$ as shown in Fig. 3(c). Rotational distortion of RuO$_6$ without tilt [see Fig. 3(d)] reduces the symmetry and gives rise to the $(\sqrt{2} \times \sqrt{2})R45°$ pattern as shown in Fig. 3(e). Forbidden fractional spots along two perpendicular directions, shown with dotted glide lines, is caused by the preserved glide symmetry. If there are both tilts and rotations of octahedra, as in Fig. 3(f), then one or the other glide symmetry should be broken, and the corresponding simulated diffraction pattern is given in Fig. 3(g). Domains with differently oriented tilts and rotations break both the glide symmetries and hence no missing fractional spots would occur.

In our LEED results from Fig. 2, we did not observe any missing spots to indicate the presence of glide symmetry. One explanation for the absence of glide line can be structure domains where both octahedral tilt and rotation exist. Our STEM study, as will be discussed below, does in fact confirm the presence of structural domains. The question remains whether the octahedral distortions are just on the single layer at the surface that minimizes the surface energy or extends throughout the thickness as a means of octahedral mismatch accommodation.

To accomplish a rigorous structural analysis of SrRuO$_3$ thin film, we performed STEM experiments to probe the atomically resolved structure. For this purpose, we choose a thicker film (22 u.c.) and cap it with STO at the top to protect the surface. Two STEM imaging modes are commonly used. High-Angle Annular Dark-Field (HAADF) imaging is more sensitive to heavy elements, and Annular Bright Field (ABF) imaging enables us to see light elements such as oxygen. Figures 4(a-b) show HAADF- and ABF- STEM images taken along [100] direction of the cubic STO, respectively. HAADF images make the quantitative analysis of the lattice parameters possible. ABF images are used to resolve the oxygen positions, which can then be used to quantify the spatially resolved oxygen octahedral distortions. Figures 4 (d-e) show the in-plane and out-of-plane lattice parameters measured from the HAADF image, respectively. The in-plane lattice parameter is equal to that of the STO bulk, while the out-of-plane lattice parameter is slightly larger than the bulk SRO value of 3.926 Å, accommodating the compressive strain from the substrate.

Figure 4(f) shows the layer-by-layer octahedral tilt evolution of the SRO films extracted from the oxygen atom displacement in the ABF image. The octahedral tilt is defined as the Ru-O-Ru bond angle ($\theta$), as shown in Fig. 4(c). The angle ($\theta$) varies with thickness across the film. The substrate is cubic and has no octahedral tilts, $\theta = 180°$. There is an intermediate region near the interface of 4-5 u.c. shown shaded in yellow in Fig. 4 (f), where the tilts are suppressed before the film is completely relaxed to 168°. It is worth noting that the tilt begins with the first u.c. of RuO$_6$ octahedra. This is consistent with the observed ($\sqrt{2} \times \sqrt{2}$)R45° LEED pattern from a single u.c. SRO film [Fig. 2(f)] where the in-plane rotational distortion is confirmed. The LEED observation is complementary to the STEM imaging. The in-plane rotation (about the $(001)_p$ axes) of octahedra, which results in the displacements of neighboring oxygen atoms along the beam direction, cannot be resolved with STEM. We also observed a reduced tilt of a few layers' octahedra at the top of the SRO film where we have capped with STO, to protect the surface. The capping with STO also suppresses the tilt distortions in agreement with earlier studies [23].

One interesting thing is to see how the lattice distortions evolve across the interface, given by a mismatch in structural symmetry between films and substrates. Both bulk SRO and CaRuO$_3$ (CRO) are orthorhombic, but CRO has larger lattice distortion and is less conductive. When interfaced with STO, they exhibit different effect on the interface structural distortions. Figure 4(g) shows the oxygen octahedral tilts near the interface of SRO and CRO with STO.

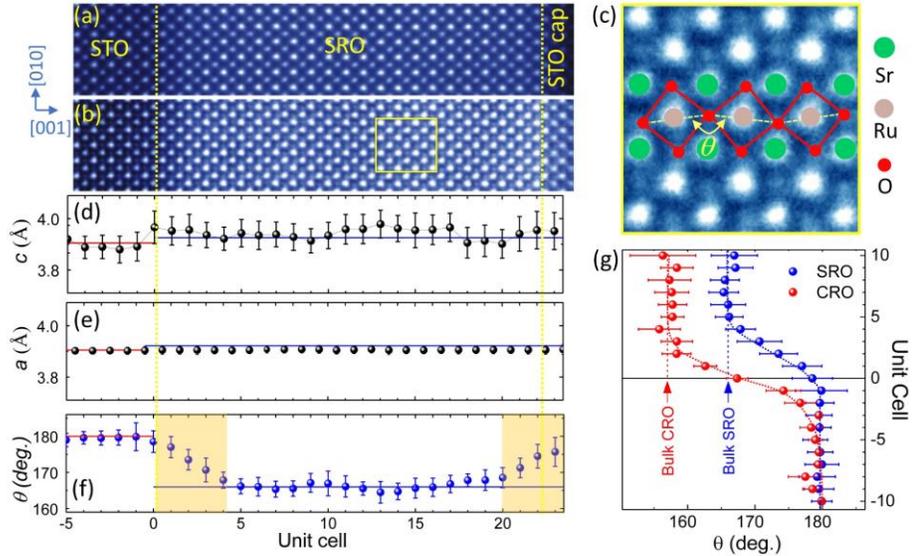

**Fig. 4** Quantitative analysis of a 22 u.c. SRO film with STEM: (a) HAADF and (b) ABF images taken along the [100] direction of the substrate. (c) A zoom-in of a region of ABF image in (b) overlapped with schematic octahedral rotations. (d) The out-of-plane (OOP) and (e) in-plane (IP) lattice parameters extracted from the HAADF images. Both the substrate STO-film interface and the capping STO-film interface are shown as orange dashed lines. Red and blue lines mark the pseudo-cubic lattice parameters of the corresponding bulk structures, respectively. (f) The oxygen octahedral tilt angles extracted from the ABF image in (b) is plotted vs. film thickness. The Ru-O-Ru bond angle of 180° for STO (red line) and 166° for bulk orthorhombic SRO (blue line) as well as the emphasized interface regions (shaded) are marked. (g) A comparison of octahedron tilts near interface of SRO/STO to CRO/STO.

Both films take about the same thickness (4-5 u.c.) from the interface to fully relax to the bulk orthorhombic structure, regardless that SRO is ~0.5% under compressive strain while CRO is ~0.6% under tensile strain. In SRO/STO, TiO$_6$ octahedra near the interface of the STO substrate are undistorted by maintaining a Ti-O-Ti bond angle of 180°. The accommodation of the interfacial rotation mismatch happens only in the RuO$_6$ octahedra of the film. In contrast, for CRO/STO, the accommodation involves tilting of the TiO$_6$ octahedra in the top few layers of the substrate. Similar tilting of the TiO$_6$ octahedra [20] as well as interfaced induced polarization of STO [19] were observed when interfaced with LSMO thin films. This shows that the octahedra in SRO are more amenable to tilt and rotational distortions than CRO and

LSMO, thus exhibiting an abrupt structural relaxation across the interface with no effect on the substrate. It is likely because SRO is less distorted in lattice and more itinerant in electronic structure than CRO, though further study is need.

SRO films compressively strained to cubic STO substrate have equal in-plane lattice parameters ($a_p = b_p$). The compressive strain causes an extension in the out-of-plane lattice parameter ($c_p$). Since the lattice parameters of the orthorhombic SRO are unequal ($a_o \neq b_o$), $c_p$ must be inclined ($\alpha_p \neq 90$), away from the $[001]_p$ axis as shown in Fig. 5(a). According to Glazer notation, the only possible tilt system is $a^+a^-c^-$ (#9) where two pseudo-cubic axes are inclined to each other and the third axis is perpendicular to both [30]. Therefore, we can expect four 90° rotated domains shown in Fig. 5(b). We define the four domains as A, B, A' and B'. Domains A and A' as well as domains B and B' are indistinguishable in STEM images. These domains are distinguished by the direction of the in-phase rotation axis (shown in red arrow) of the octahedra with respect to the cubic axes of the substrate. Domains A and A' have the in-phase rotation axis along the [100] of the substrate while domains B and B' have the in-phase rotation axis along the [010] direction.

STEM images taken along the [100]p direction show two structural domains (A/A' vs. B/B'). In addition to the structure shown in Fig. 4, we also observed another region with a different structural symmetry. The two structures are clearly distinguishable by comparing the Fast Fourier Transform (FFT) results of the ABF images. Figure 5(c) shows the FFT of the ABF image of the structural domain shown in Fig. 4 and the simulated diffraction pattern obtained using orthorhombic SRO ($a^+a^-c^-$ model structure) with the in-phase rotation axis as the beam direction. Simply based on the pattern

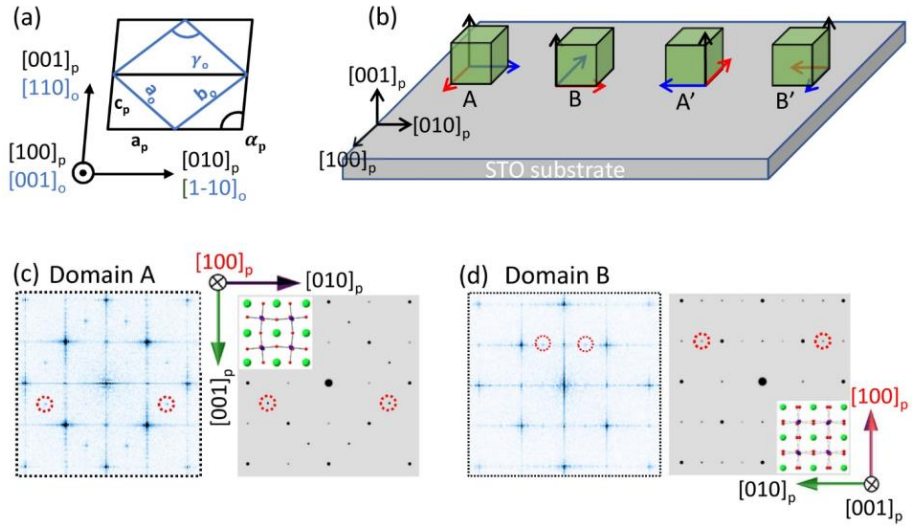

**Fig. 5** (a) Distorted orthorhombic film on cubic substrate under compressive strain. 'o' and 'p' index refer to the orthorhombic and pseudo-cubic lattice parameters, respectively. (b) Four 90-degree rotated domains are expected as shown. Red arrow points to the direction of in-phase rotation of the neighboring octahedra in Glazer notation. Two distinctly different domains are observed in STEM taken along [100] beam direction. (c) FFT of domain A and simulated diffraction obtained with the beam direction the same as the direction of in-phase rotation of the model structure (inset). (d) FFT of domain B and simulated diffraction with the beam direction perpendicular to the in-phase rotation axis.

symmetry without detailed quantitative analysis of octahedral rotations, the simulated pattern agrees well with the observed FFT image. We therefore identify this structure to be that of the domain A (and/or A') shown in Fig. 5(b). The in-phase rotation axis is shown in red and is along the [100] direction of the STO substrate.

The experimentally observed FFT pattern shown in Fig. 5(d) (left) for the second structural domain has a different symmetry, as compared with the FFT pattern for domain (A/A') shown in Fig. 5(c). The experimental FFT matches the simulated diffraction pattern of the 90° rotated structure of the previously described $a^+a^-c^-$ model structure, which is shown in the right panel of Fig. 5(d) with additional (1 × 2) diffraction spots (red circled). This structural domain is thus identified as domain B (B') as shown in Fig. 5(b). In principle, LEED could display such a (1 × 2) diffraction pattern if electron mean free path was long enough to reach the second unit cell from the surface to provide the information. Nevertheless, two types of structural domains observed in the TEM are identified as the two 90° rotated structural domains of the $a^+a^-c^-$ structure of SRO.

In summary, we demonstrate that octahedral rotational and tilt distortions appear in ultrathin SRO films, down to single u.c. thickness, even when grown on a cubic perovskite substrate like STO. Such distortions reduce the symmetry of the surface and thus give rise to fractional diffraction spots in the ($\sqrt{2} \times \sqrt{2}$)R45° LEED pattern. Yet it takes 4 - 5 unit cells from the interface before SRO film fully relaxes back to its bulk orthorhombic structure, as observed in STEM measurements. The presence of 90° in-plane rotated structural domains is also established. On the other hand, the octahedral rotational and tilt distortions in SRO thin films do not affect the interface layer structure of the STO substrate, in contrast to CRO and LSMO thin films. The presence of these octahedral distortions in ultrathin films are important to understand the different transport and magnetic properties, such as the thickness dependent metal-insulator transition and the loss of ferromagnetism [31–34].

*Acknowledgments*: This work was primarily supported by the U.S. DOE under Grant No. DOE DE-SC0002136. The STEM work done at Brookhaven National Laboratory was sponsored by the US DOE Basic Energy Sciences, Materials Sciences and Engineering Division under Contract No. DESC0012704. P.A. was partially supported by the US NSF under Grant No. DMR 1608865.